\documentclass[aps,pra,reprint,showpacs,floatfix,amsmath,amssymb,10pt]{revtex4-1} 

\usepackage[pdftex]{graphicx}       
\usepackage{float}

\usepackage{epstopdf}
 \usepackage{txfonts}        

\usepackage{amsmath}				

\newcommand{\dd}{\mathrm{d}}
\newcommand{\pd}{\partial}
\newcommand{\hb}{\hbar}

\begin{document}


\title{Universality and itinerant ferromagnetism in rotating strongly interacting Fermi gases}

\author{B. C. Mulkerin} 
\author{C. J. Bradly} 
\author{H. M. Quiney}
\author{A. M. Martin}
\affiliation{School of Physics, University of Melbourne, Victoria 3010, Australia}

\date{\today}
\begin{abstract}
We analytically determine the properties of three interacting fermions in a harmonic trap subject to an external rotation. Thermodynamic quantities such as the entropy and energy are calculated from the third order quantum virial expansion. By parameterizing the solutions in the rotating frame we find that the energy and entropy are universal for all rotations in the strongly interacting regime. Additionally, we find that rotation suppresses the onset of itinerant ferromagnetism in strongly interacting repulsive three-body systems.
\end{abstract}

\maketitle



Broad Feshbach resonances in two-component atomic Fermi gases have made it possible to explore the crossover between Bardeen-Cooper-Schrieffer (BCS) superfluidity and Bose-Einstein condensation (BEC) \cite{Chin2010,Shin:2008,zwierlein-2006-442}. In the strongly interacting regime the difficulties in developing many-body theories for these systems has motivated the study of exact solutions to few-body problems as a means to gain insight into the many-body problem \cite{Busch:1998,Werner2006,Werner2006a,Kestner2007,Daily2010,Liu2009,Liu2010,Daily2010,Staferle2006}. Of particular interest has been the prediction \cite{Liu2010,Rakshit2012,Kaplan2011,Bhaduri2012} and observation \cite{Nascimbene2010,Ku2012} of universality in the strongly interacting regime of thermodynamic quantities, such as the energy and entropy. To date such calculations have been restricted to gases in non-rotating spherically symmetric harmonic traps.

In this work we consider the properties of a strongly interacting gas in a rotating trap. In particular we solve the two- \cite{Mulkerin2012} and three-body problems and calculate the virial expansion of the thermodynamic potential to third order, enabling the calculation of thermodynamic quantities. Interestingly, we find in the thermodynamic limit that the second- and third-order virial coefficients are universal, with respect to an external rotation and trapping frequencies. From this we show that thermodynamic quantities such as the energy and entropy are universal with respect to rotation through a simple rescaling of the Fermi energy.

In addition to these thermodynamic results we also examine the interplay between rotation and the emergence of itinerant ferromagnetism in strongly interacting ultra-cold Fermi gases. In the original work of Stoner \cite{stoner1938} it was proposed via a meanfield theory that a repulsive Fermi gas will always exhibit a ferromagnetic phase. Most recent experimental evidence \cite{Sanner2012} suggests there is no transition. Current theoretical work, Monte Carlo simulations and Tan relations \cite{Liu2010,Chang2011,Conduit2009a,Conduit2009} are contradictory. In this work we show that for the three-body problem, rotation suppresses the emergence of itinerant ferromagnetism. Additionally, in the thermodynamic limit itinerant ferromagnetism is suppressed for temperatures $T>10^{-7}T_{\text{F}}$ as the rotation frequency approaches the trapping frequency.



Our starting point is the wavefunction, $\psi(\mathbf{r}_1,\mathbf{r}_2,\ldots)$, of $N$ particles interacting in the $s$-wave channel at low energies. This satisfies the Bethe-Peierls boundary conditions 
\begin{alignat}{1}
\lim_{r_{ij}\rightarrow 0}\partial(r_{ij}\psi)/\partial r_{ij} = -r_{ij} \psi  /a 
\label{eq:bethe-peierls},
\end{alignat}
where the interaction is parametrized by the scattering length $a$, and $r_{ij}=|\mathbf{r}_i-\mathbf{r}_j|$ is the separation of opposite-spin fermions.
Away from $r_{ij}=0$, the wavefunction of $N$ particles in a spherically symmetric rotating harmonic trap satisfies the non-interacting Schr\"odinger equation, 
\begin{alignat}{1} \label{eq:rotHamiltonian}
\sum_{i=1}^{N} \left [ -\frac{\hbar^2}{2\mu}\mathbf{\nabla}_{i}^2 + \frac{1}{2}\mu\omega^2\mathbf{r}_i^2+ i\hbar\Omega_z\partial_{\phi_i}\right ] \psi=E\psi,
\end{alignat}
where $\mathbf{r}_i$ and $\mu$ are the position and mass of each particle and $\omega$ and $\Omega_z$ are the trapping and rotation frequencies, the latter assumed to be defined about the $z$ axis. 

The center-of-mass Hamiltonian can be decoupled from Eq.~\eqref{eq:rotHamiltonian} and defines the rotating harmonic motion of a particle of mass $M=N\mu$ with energy spectrum 
\begin{alignat}{1}
E_{\text{cm}}=(2n+l+3/2)\hbar\omega + m\hbar\Omega_z,
\label{eq:SingleParticleEnergies}  
\end{alignat}
where $n,l,m$ label the usual harmonic oscillator eigenstates $R_{nl}Y_l^m$. The relative energy,  $E_\text{rel}=E-E_\text{c.m.}$, incorporates the effects of the contact interaction but not the effects of the external rotation, since only $s$-wave states may interact. For two opposite-spin fermions the wavefunction in relative coordinates that satisfies the Bethe-Peierls boundary condition Eq.~\eqref{eq:bethe-peierls} is \cite{Busch:1998}
\begin{alignat}{1}
\psi^{\text{rel}}_{2b}(\mathbf{r};\nu)\propto\Gamma\left(-\nu\right)U\left(-\nu,3/2,r^2/d^2\right)\,\text{exp}\left(-r^2/d^2\right),
\end{alignat}
where $U$ is the confluent hypergeometric function of the second kind. A pseudo-quantum number, $\nu$, parametrizes the relative energy $E_\text{rel}=(2\nu+3/2)\hbar\omega$, and satisfies the relation
\begin{equation}\label{eq:Transcendental}
\frac{2\Gamma(-\nu)}{\Gamma(-\nu-1/2)}=\frac{d}{a},
\end{equation}
for harmonic oscillator length $d=\sqrt{\hbar/(\mu\omega)}$. In particular, for the unitary limit, where $a\rightarrow\pm\infty$, the relative energy spectrum simplifies to $E_\text{rel}=(2n+1/2)\hbar\omega$, where $n$ is any no-negative integer.


To find the energy spectrum of three interacting fermions in a rotating trap, we consider the configuration of two spin up fermions and one spin down, $\uparrow\downarrow\uparrow$, where two opposite spin particles interact at a point and form a pair, and the third moves relative to the pair. We define the center-of-mass coordinate of the three particles as $\smash{\mathbf{R}=(\mathbf{r}_1+\mathbf{r}_2+\mathbf{r}_3)/3}$, the relative coordinate between the interacting pair, $\smash{\mathbf{r}=\mathbf{r}_1-\mathbf{r}_2}$ and the relative coordinate between the third non-interacting particle and the center-of-mass of the pair as $\smash{\mathbf{\rho}=(2/\sqrt{3})[\mathbf{r}_3-(\mathbf{r}_1+\mathbf{r}_2)/2]}$. In this Jacobi coordinate system the center-of-mass Hamiltonian 
decouples from the relative Hamiltonian
\begin{alignat}{1}
H_{\text{rel}}= 
- \frac{\hbar^2}{\mu} \left(\nabla^{2}_{\mathbf{r}} + \nabla^{2}_{\mathbf{\rho}}\right) + \frac{1}{4}\mu\omega^2(\mathbf{r}^2 + \mathbf{\rho}^2) - i\hbar\Omega_z\partial_{\phi_\rho} \label{eq:jacobiHamil},
\end{alignat}
where $\mu/2$ is the reduced mass of the interacting pair. Like the two-body system, the angular momentum vanishes in the interacting pair but the third fermion can rotate around the pair and be affected by the external rotation, $\Omega_z$. This couples higher order angular momentum states to lower energies in the system.
%
In order to solve the relative Hamiltonian~\eqref{eq:jacobiHamil} we take  
\begin{alignat}{1}
\Psi_{3b}^{\text{rel}}(\mathbf{r},\mathbf{\rho})=(1-P_{13})\sum_{n=0}^\infty c_n \psi_{2b}^{\text{rel}}(\mathbf{r};\nu_{nlm})R_{nm}(\mathbf{\rho})Y_{l}^{m}(\hat{\rho}),
\label{eq:ansatz}
\end{alignat}
as an ansatz for the wavefunction, where $P_{13}$ is an operator that exchanges the spin $\uparrow$ particles. 
The eigenenergies of this system are
\begin{alignat}{1}
E_{\text{rel}}=\left[\left(2n+l+\frac{3}{2}\right)+\left(2\nu_{nlm}+\frac{3}{2}\right)\right]\hbar\omega+m\hbar\Omega_z \label{eq:3energies}.
\end{alignat}
The presence of the rotational term, $m\hbar\Omega_z$, in the eigenenergy spectrum shifts the non-rotating energy spectrum found in \cite{Liu2009}. To solve for the coefficients $c_n$ in Eq.~\eqref{eq:ansatz} we use the Bethe-Peierls boundary condition Eq.~\eqref{eq:bethe-peierls} and choose a set of quantum numbers $nlm$ and energy $E_{\text{rel}}$ to solve for a particular $\nu_{nlm}$ and scattering length $a$ \cite{Liu2009}. 


\begin{figure}
	\includegraphics{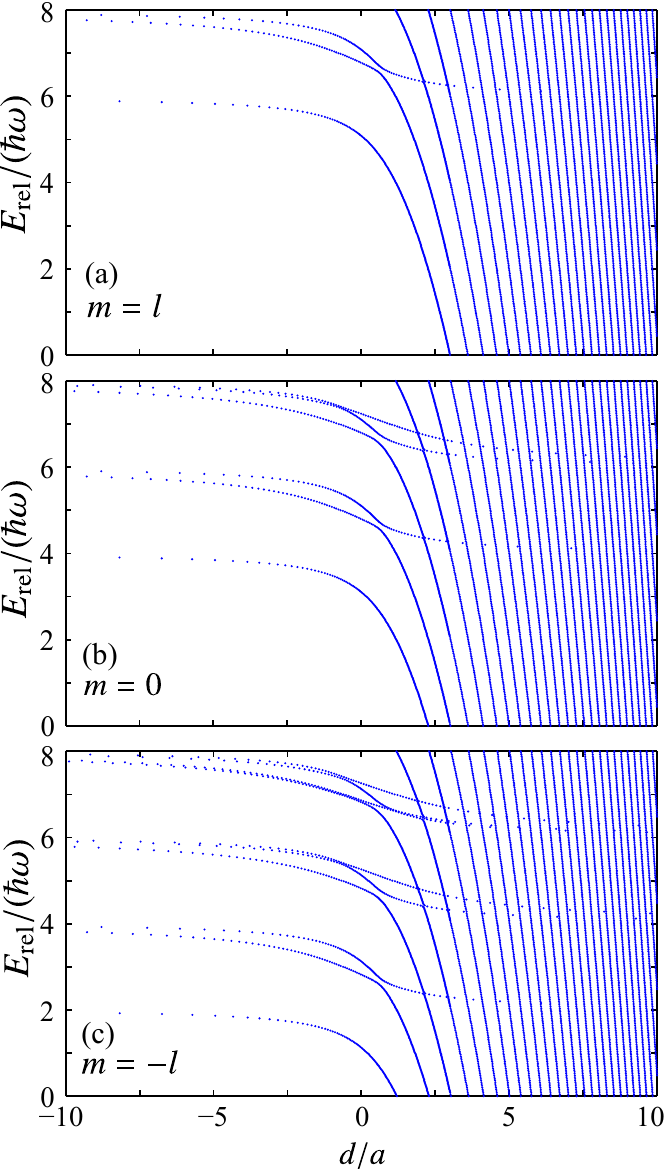}
	\caption{Energy spectrum of three interacting fermions with $l=2$ and $m=-2,0$ and $2$ for plots (a), (b) and (c) respectively with a rotation of $\Omega_z=0.9\omega$. We can see the shifting of the energy spectrum with the rotation included, in particular we can see the number of lower energy states increase for $m=-2$.}
	\label{fig:spectrum}
\end{figure}	

The relative energy spectrum can be found numerically for any scattering length, but in the unitary regime it is simpler to use the method of Werner and Castin \cite{Werner2006a} to obtain the energies using hyperspherical coordinates $(R,\alpha,\hat{r},\hat{\rho})$, where $R=\sqrt{(r^2+\rho^2)/2}$ is the hyperradius, $\alpha=\text{arctan}(r/\rho)$ is the first hyperangle and $\hat{r}$ and $\hat{\rho}$ are the direction of each Jacobi coordinate. Using this coordinate system and the ansatz for the wavefunction \cite{Werner2006a},
\begin{alignat}{1}
\Psi_{3b}^{\text{rel}}(\mathbf{r},\mathbf{\rho})=\frac{F(R)}{R^2}(1-P_{13})\frac{\varphi(\alpha)}{\sin(2\alpha)}Y_{l}^{m}(\hat{\rho}),
\label{eq:Fadeev}
\end{alignat}
the Hamiltonian, Eq.~\eqref{eq:jacobiHamil}, can be written as two decoupled Schr\"odinger equations,
\begin{alignat}{1}
-\frac{\hbar^2}{2m}\left(F''+\frac{1}{R}F'\right)+\left ( \frac{\hbar^2 s^{2}_{nl}}{2m R^2}+\frac{1}{2}m\omega^2 R^2 -m\hbar\Omega_z \right )F =E_{\text{rel}}F \label{eq:hyperradial},
\end{alignat}
and
\begin{alignat}{1}
-\varphi''(\alpha)+\frac{l(l+1)}{\cos^2(\alpha)}\varphi(\alpha)=s_{nl}^2\varphi(\alpha) \label{eq:hyperangular}.
\end{alignat}
For three fermions $s_{nl}^2$ is always positive and we can interpret the hyperradial Sch\"odinger equation \eqref{eq:hyperradial} as a particle moving in a two dimensional effective potential $\bigl( \hbar^2 s^{2}_{nl}/(2 m R^2) + \tfrac{1}{2} m \omega^2 R^2 \bigl)$ with energy spectrum \cite{Werner2006}
\begin{alignat}{1}
E_{\text{rel}}=(2q + s_{nl} + 1)\hbar\omega + m\hbar\Omega_z \label{eq:unitary},
\end{alignat}
where $q$ is a positive integer. The solutions to Eq.~\eqref{eq:hyperangular} must satisfy $\varphi(\pi/2)=0$ so that the ansatz Eq.~\eqref{eq:Fadeev} does not diverge. Hence,
\begin{alignat}{1}
\varphi(\alpha)\propto \cos(\alpha)^{l + 1}P^{(l + 1/2,-1/2)}_{(s_{nl} - l - 1)/2} \big [ -\cos(\alpha) \big ],
\label{eq:PhiExpansion}
\end{alignat}
where $P^{(\gamma,\beta)}_{n}(x)$ is the regular Jacobi polynomial  \cite{abramowitz+stegun}. The eigenvalues $s_{nl}$ are determined from the Bethe-Peierls boundary condition \eqref{eq:bethe-peierls}, which 
in hyperspherical coordinates reads
\begin{alignat}{1}
\varphi'(0)-(-1)^l \frac{4}{\sqrt{3}}\varphi \left ( \frac{\pi}{3} \right) =0 
\label{eq:strongBethe}.
\end{alignat}
The procedure for solving Eq.~\eqref{eq:strongBethe} for values of $s_{nl}$ using the general solutions to the hyperangle equation is given in \cite{Liu2010}. It can be shown that the two spectra Eq.~\eqref{eq:unitary} and \eqref{eq:3energies} are the same in the unitary limit, where the solutions from Eq.~\eqref{eq:3energies} are found numerically. 


\begin{figure}[t!]
	\includegraphics{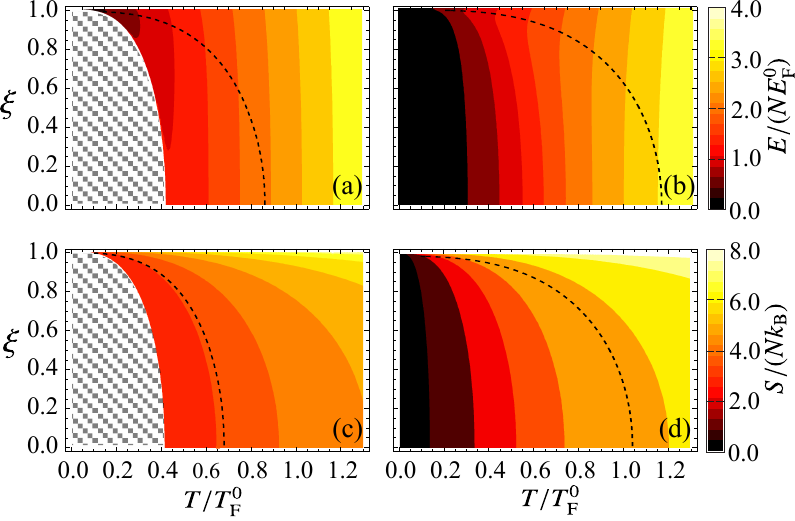}
	\caption{
(a,b), Energy per particle $E/(NE_{\text{F}}^0)$ and (c,d), entropy per particle $S/(Nk_{\text{B}})$ for dimensionless rotation $\xi$ as a function of reduced temperature $T/T_{\text{F}}^0$, in the strongly (a,c), attractive and (b,d), repulsive regimes. For comparison we plot the temperature at which the second and third order virial expansions differ by $1\%$ for a rotation $\xi$, the dashed line. The hashed region in plots (a) and (c) indicate an unphysical solution to the virial expansion.}
	\label{fig:contours}
\end{figure}	

The energy spectrum of three interacting fermions is shown in Fig.~\eqref{fig:spectrum} for a rotation of $\Omega_z=0.9\omega$, a relative angular momentum of $l=2$ and axial angular momentum of $m=-2,0,2$. We see the energy levels for the three interacting fermions shift for each $m$ quantum number, lifting the $(2l+1)$ fold degeneracy.

To obtain the repulsive spectrum we omit the solution energy levels which are the lowest in each $l$-subspace lowest order $n=0$ bound state in each $l$ subspace, i.e. the $s_{0l}$ energy levels. The lowest energy in the repulsive regime is then the relative energy, which is dependent upon the rotation $\Omega_z$, plus the center-of-mass energy,
\begin{alignat}{1}
E_{gs}^{\uparrow\downarrow\uparrow}=(s_{nl} + 1)\hbar\omega + m\hbar\Omega_z + 1.5\hbar\omega.
\label{eq:intground}
\end{alignat}
This can be compared to the energy of three polarized fermions, which is the sum of the three lowest allowed energies:
\begin{alignat}{1}
E_{gs}^{\uparrow\uparrow\uparrow}=1.5\hbar\omega +2.5\hbar\omega + [1.5\hbar\omega+(\omega-\Omega_z)\hbar].
\label{eq:nonintground}
\end{alignat}
Comparing Eqs.~\eqref{eq:intground} and \eqref{eq:nonintground} it is possible to find a critical rotation, $\Omega_c$, for which the ground state energy of three repulsively interacting fermions becomes lower than three non-interacting polarized fermions. This is a rotation for which the three non-interacting fermions are unstable with respect to the three interacting fermion system. Choosing the lowest relative energy level in the $\smash{l=2}$ subspace, $\smash{s_{12}\simeq4.80}$, and a magnetic quantum number $\smash{m=-l}$, we find a critical rotation of $\smash{\Omega_c\simeq 0.8\omega}$. Since the ground state energy of three repulsively interacting fermions can be controlled by varying an external rotation, this indicates that an itinerant ferromagnetic transition, in the three-body system, only occurs for a rotation $\Omega_z<\Omega_c$. 



The two- \cite{Mulkerin2012} and three-body solutions can now be used to calculate the many-body properties of strongly interacting rotating Fermi gases. This is done via a quantum virial expansion of the grand thermodynamic potential, $\Phi=-k_B T \ln \mathcal{Z}$, in terms of the fugacity $z$: 
\begin{alignat}{1}
 \Phi= & -k_B T Q_1 \frac{1}{2} \int_0^\infty \!\!\!\! \dd \epsilon \epsilon^2 \ln(1+z e^{-\epsilon}) \nonumber \\ & - k_B T Q_1 \left(z+\Delta b_2 z^2+\Delta b_3 z^3+\dots\right),
\label{eq:GTPOmega}
\end{alignat}
where
%
\begin{alignat}{1}
& \Delta b_2= \Delta Q_2/Q_1, \label{eq:Deltab2} \\ &
\Delta b_3= \Delta Q_3/Q_1 - \Delta Q_2, \label{eq:Deltab3}
\end{alignat}
and $\smash{\Delta Q_n = Q_n - Q_n^{(0)}}$ with $\smash{Q_N=\text{Tr}\left[\exp(-\mathcal{H}_N/k_B T)\right]}$ \cite{HuangBook}.
To first order for dimensionless rotation $\xi=\Omega_z/\omega$, $Q_1$ is given by
\begin{equation} 
Q_1 = \left(\frac{k_{\text{B}}T}{\hbar\omega}\right)^3\frac{1}{1-\xi^2} +\dots
\label{eq:Q1}
\end{equation}
Following \cite{Mulkerin2012}, the second-order virial coefficient for a rotating trapped gas in the high temperature limit is
\begin{alignat}{1}
\Delta b_2^\text{att} & = \frac{1}{4} - \frac{\tilde{\omega}^2}{32}  + \dots \label{eq:secondatt}\\ 
\Delta b_2^\text{rep} & = -\frac{1}{4} + \frac{\tilde{\omega}}{4}  + \dots \label{eq:secondrep}
\end{alignat}
where $\tilde{\omega}=\hb\omega/k_{\text{B}} T$ is the reduced trapping frequency and $\smash{\tilde{\omega} \to 0}$ represents the thermodynamic limit. Extending the work of non-rotating systems in \cite{Liu2009} we find that $\Delta b_3^{att}$ is universal for any rotation and given by
\begin{alignat}{1}
\Delta b_3^\text{att} \simeq  -0.06833960 + O(\tilde{\omega}^2). \label{eq:thirddatt}
\end{alignat}
For a repulsive gas the third-order virial coefficient to lowest order in $\tilde{\omega}$ is also universal,
\begin{alignat}{1}
\Delta b_3^\text{rep} \simeq  0.34976 +O(\tilde{\omega}). \label{eq:thirdrep}
\end{alignat}
Despite the rotational dependence of the two- and three-body eigenspectrums the second- and third-order virial coefficients are independent of rotation in the thermodynamic limit.

We are now able to calculate the total energy $E=-3\Phi$ and entropy $\smash{S=-\partial\Phi/\partial T}$ from the thermodynamic potential $\Phi$ of a strongly interacting gas \cite{HuangBook}. To determine the thermodynamic potential, the fugacity $\smash{z=\text{exp}(\mu/k_{\text{B}}T)}$ must be calculated from the total number of particles $\smash{N=-\pd\Phi/\pd\mu}$, where $\mu$ is the chemical potential. 


In Fig.~\ref{fig:contours} we plot the energy, in units of $\smash{NE_{\text{F}}^0}$, where $\smash{E_{\text{F}}^0=(3N)^{1/3}\hbar\omega}$ is the non-rotating Fermi energy, and entropy per particle as functions of reduced temperature, $\smash{T_{\text{F}}^0=E_{\text{F}}^0/ k_{\text{B}}}$, and rotation using the virial expansion to third-order for a strongly repulsive, (b,d) and attractive, (a,c) Fermi gas. The hashed areas in plots (a) and (c) correspond to solutions of the energy and entropy that are unphysical. The dashed curve in Fig.~\ref{fig:contours} is the temperature at which the second and third order expansion differ by $1\%$, providing a conservative estimate of the temperature range of validity for the virial expansion. As the rotation is changed, for a fixed temperature, the energy (a,b) and entropy (c,d) change. Hence, the thermodynamic quantities appear not to be universal with respect to rotation.


From the three-body calculations Eqs.~\eqref{eq:intground} and \eqref{eq:nonintground} we see that itinerant ferromagnetism is suppressed for $\smash{\Omega_z\gtrsim0.8\omega}$. Fig.~\ref{fig:contours}(b) plots the total energy in the strongly repulsive regime. As $\smash{\xi \to 1}$ the validity of the solutions extends to $\smash{T \to 0}$. In this regime, by comparing the energy of the strongly interacting gas with the equivalent non-interacting polarized gas, we find that the itinerant ferromagnetic phase is suppressed for $T/T_{\text{F}}^0=10^{-7}$ \footnote{Due to numerical instabilities it is not possible to set $\xi=1$. Hence the limit $\smash{\xi\to 1}$ is evaluated at $\smash{\xi=1-10^{-7}}$, for which the virial expansion is valid for $\smash{T/T_{\text{F}}^0}>10^{-7}$}.



\begin{figure}[t]
	\includegraphics{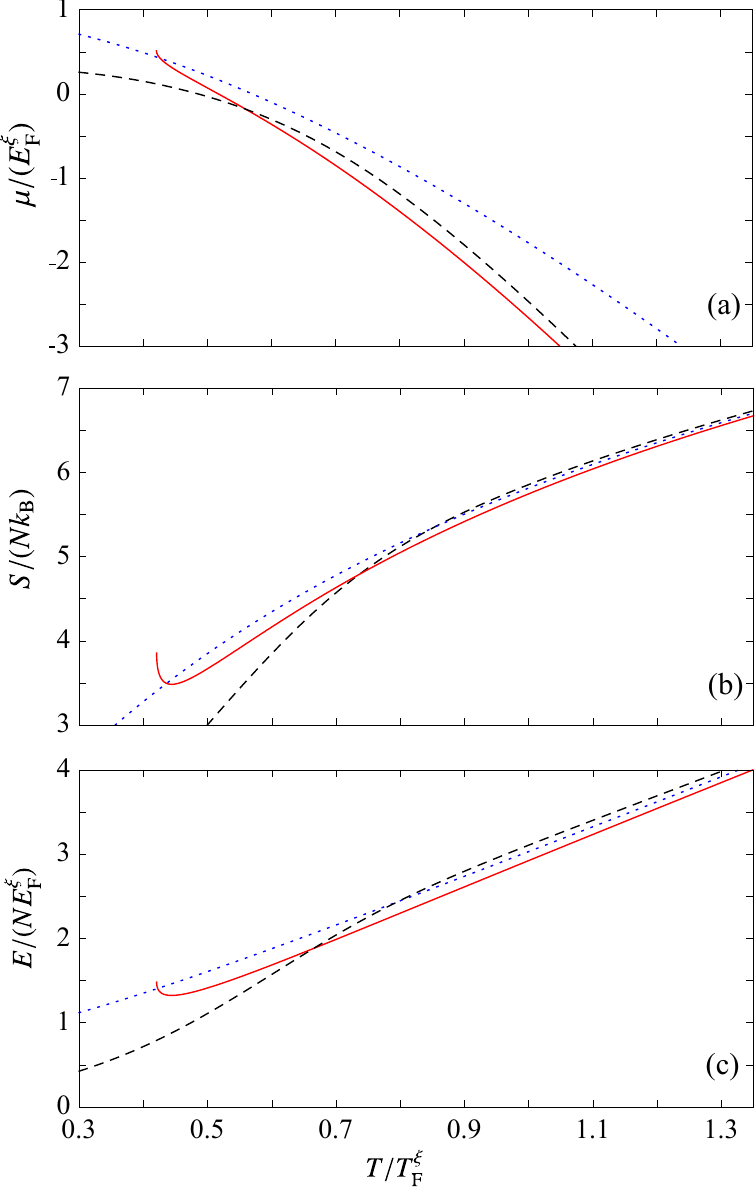}
	\caption{Chemical potential, (a), entropy per particle, (b) and energy per particle, (c) of a strongly attractive (solid), repulsive (dashed) and ideal Fermi gas (dotted) for dimensionless rotation $\xi=0, 0.5, 0.9$ and $0.99$ as a function of reduced temperature $\smash{T/T_{\text{F}}^\xi}$. }
	\label{fig:energyentropy}
\end{figure}

Figure~\ref{fig:contours} demonstrates that for a given rotation frequency, $\xi$, there is a universal dependence of $\smash{E/(NE_{\text{F}}^{0})}$ and $\smash{S/(Nk_{\text{B}})}$ as a function of $T_{\text{F}}^{0}$, with respect to particle number and trapping frequency. However, through a simple rescaling of the Fermi energy [temperature] of the form $\smash{E_{\text{F}}^{\xi}=E_{\text{F}}^{0}(1-\xi^2)^{-1/3}}$ $\smash{\bigl[T_{\text{F}}^{\xi}=T_{\text{F}}^{0}(1-\xi^2)^{-1/3}\bigl]}$ it is possible to remove the rotational dependence observed in Fig.~\ref{fig:contours}. This generalized universality arises from the fact that under this rescaling the functional dependence of $\xi$ is removed from both the single particle cluster function, $Q_1$ [Eq.~\eqref{eq:Q1}], and the chemical potential. To emphasize the universal nature of the chemical potential with respect to rotation, $\mu/E_{\text{F}}^{\xi}$ is plotted in Fig.~\ref{fig:energyentropy}(a) as a function of $\smash{T/T_{\text{F}}^{\xi}}$ for various rotations, $\xi$, in the strongly attractive (solid), repulsive (dashed) and ideal (dotted) regimes. Hence, in conjunction with the fact that the second- [Eqs.(\ref{eq:secondatt},\ref{eq:secondrep})] and third-order [Eqs.(\ref{eq:thirddatt},\ref{eq:thirdrep})] virial coefficients are independent of rotation in the thermodynamic limit, the thermodynamic potential is independent of rotation. As a direct consequence the rescaled energy, $\smash{E/(NE_{\text{F}}^{\xi})}$, and entropy, $\smash{S/(Nk_{\text{B}})}$,  are universal with respect to rotation as functions of $\smash{T/T_{\text{F}}^{\xi}}$,. This property is demonstrated in Figs.~\ref{fig:energyentropy}(b,c) which plot $\smash{E/(NE_{\text{F}}^{\xi})}$ (b) and $\smash{S/(Nk_{\text{B}})}$ (c) as a function of $\smash{T/T_{\text{F}}^{\xi}}$ for various values of $\xi$ in the strongly attractive (solid), repulsive (dashed) and ideal (dotted) regimes.

In conclusion we have examined the problem of three ultracold fermions in a harmonic trap subject to an external rotation. For the three-body problem we have demonstrated that rotation suppresses the transition to a ferromagnetic state. Explicitly we have shown that the three-body ferromagnetic state has a higher energy than the strongly interacting repulsive state for $\smash{\Omega_z\gtrsim0.8\omega}$ and is, consequently, unstable. Additionally, from the three-body solutions and the use of previous two-body results \cite{Mulkerin2012} we have calculated the equations of state using the virial expansion to third-order. Despite the rotational dependence of the two- and three-body eigenspectrums we have generalized the universal nature of strongly interacting fermions to include rotation by a simple rescaling of the Fermi energy and temperature. These results could be used as benchmarks in experiment to test the universal properties of strongly interacting rotating ultracold Fermi gases. \linebreak

H.M.Q. gratefully acknowledges the support of the ARC Centre of Excellence for Coherent X-ray Science.

\bibliography{3particles_3} 

\end{document}